\documentclass[epsf,twocolumn,showpacs,preprintnumbers]{revtex4}

\usepackage{amsmath,amssymb,amsfonts}
\usepackage{graphics}
\usepackage{dcolumn} 
\usepackage{bm,graphicx,color}
\usepackage{epsfig}
\newcommand{\ppd}{PbPd$_3$}
\newcommand{\spd}{SnPd$_3$}

\begin{document}
\title{Coexistence of Triple Nodal Points, Nodal Links, and Unusual Flat Bands\\ 
in intermetallic ${\cal A}$Pd$_3$ (${\cal A}$=Pb, Sn)
}
\author{Kyo-Hoon Ahn$^1$}
\author{Warren E. Pickett$^2$}
\email{pickett@physics.ucdavis.edu}
\author{Kwan-Woo Lee$^{1,3}$}
\email{mckwan@korea.ac.kr}
\affiliation{
 $^1$Department of Applied Physics, Graduate School, Korea University, Sejong 30019, Korea\\
 $^2$Department of Physics, University of California, Davis, California 95616, USA\\
 $^3$Division of Display and Semiconductor Physics, Korea University, Sejong 30019, Korea
}
\date{\today}
\begin{abstract}
We investigate the electronic structure and several properties, and topological character,
of the cubic time-reversal invariant intermetallic compounds PbPd$_3$ and SnPd$_3$
using density functional theory based methods.
These compounds have a dispersionless band along the $\Gamma-X$ line, forming 
the top of the Pd $4d$ bands and 
lying within a few meV of the Fermi level $E_F$. 
Effects of the flat band on transport and optical properties have been inspected
by varying the doping concentration treated with the virtual crystal approximation for
substitution on the Pb site.
In the absence of spin-orbit coupling (SOC), 
we find triple nodal points and three-dimensional nodal loops,
which are known to lead to surface bands and 
drumhead states, respectively, which we discuss for PbPd$_3$. 
SOC removes degeneracy in most of the zone, providing a topological index $Z_2$=1 
on the $k_z=0$ plane that indicates a topological character on that plane.
The isovalent and isostructural compound SnPd$_3$ shows only minor differences in
its electronic structures, so it is expected to display similar electronic, transport,
and topological properties.
\end{abstract}
\maketitle

\section{Introduction}
For the last decade one of the most active issues in condensed matter physics is
topological character of the electronic structure and the new properties that may
arise. Various types of topological matters have been proposed and some have
received experimental support.  In addition to topological insulators, 
three-dimensional topological semimetals  
have attracted large interest due to their unusual boundary states,\cite{armi}
which may stimulate novel directions in electronics and spintronics separate from
topological insulators.
Recently, a novel triple point fermionic phase, having no high energy counterparts,
has been also proposed in both nonsymmorphic\cite{brad16} and symmorphic structures.\cite{zhu16}
This phase, predicted along the symmetry line with 3-fold rotation and mirror ($C_{3v}$) symmetries 
in a few systems of WC-type or half-Heusler structures,\cite{hweng16a,hweng16b,yang17,wang17} 
is expected to have various unconventional properties
and is experimentally observed in MoP.\cite{ding17}


The intermetallic palladium-lead phase diagram includes various compounds.\cite{phase1,phase2}
Among them, the so-called ideal zvyagintsevite compound \ppd~ (cubic Cu$_3$Au structure)
has been investigated 
as a catalyst for electrochemical oxygen reduction.\cite{kim,disalvo16,gunji} 
It has been known that its sister compounds can absorb a large amount 
of hydrogen,\cite{ke_the,kohl} making them of interest as candidates for hydrogen storage 
or membrane separation.

Separately, flat bands in part or all of the Brillouin zone (BZ) have piqued interest for a variety of reasons.
A weakly dispersive band leads to a peak in density of states (DOS), typically much stronger
and perhaps narrower than structures arising from van Hove singularities.
When a DOS peak is close to, or at, the Fermi level $E_F$, instabilities of the simple
Fermi liquid states are encouraged; peaks can enhance 
a superconducting critical temperature or induce Stoner-type magnetic 
instability.\cite{dag94,imada00,mgcni3}
In PbPd$_3$ a remarkably flat band appears along the $\Gamma-X$ lines, 
arising from a lack of $dd\delta$ hopping such as occurs in (cubic) perovskite-like 
systems.\cite{mgcni3,LP09}
Because of the limited phase space where the band is flat, it leads to two-dimensional-like
step in the DOS (see below).
It becomes of importance because it lies within a few meV of $E_F$, 
implying that transport, thermodynamic, and infrared properties will display unusual dependences
on temperature, doping, or other changes in the system (viz. strain).

The manuscript is organized as follows. Sec. II describes our calculational methods 
based on an {\it ab initio} approach. The electronic structure and topological 
characters of \ppd~ are presented in Sec. III.
Section IV discusses the effects of the unusual flat band on transport and optical properties.
A brief comparison of electronic structures between \ppd~ and \spd~ is addressed in Sec. V.
Finally, Sec. VI provides a summary.

\begin{figure}[tbp]
{\resizebox{7.5cm}{5cm}{\includegraphics{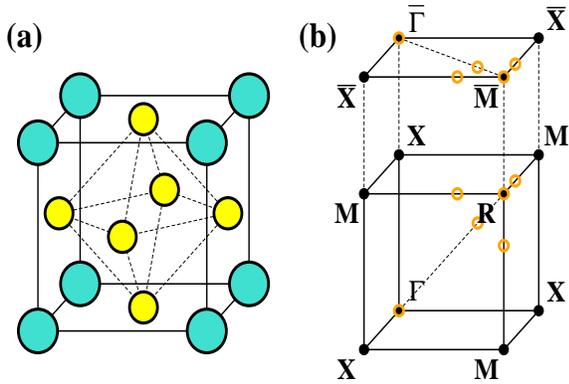}}}
\caption{(a) Cubic $L2_1$ crystal structure of ${\cal A}$Pd$_3$,
having a Pd$_6$ octahedron.  
(b) The bulk and (001) surface Brillouin zones (BZs) with high symmetry points.
The unfilled circles denote positions of spinless triple nodal points,
appearing in the range of --0.3 eV to 0.2 eV.
}
\label{str}
\end{figure}

\section{Approach}
We have performed density functional theory calculations 
based on the exchange-correlation functional of the Perdew-Burke-Ernzerhof 
generalized gradient approximation (GGA)\cite{gga} 
with the all-electron full-potential code {\sc wien2k}.\cite{wien2k}
Since Pb is a heavy atom, spin-orbit coupling (SOC) is included in all calculations unless otherwise noted.
Figure \ref{str} (a) shows the cubic Cu$_3$Au crystal structure 
(space group: $Pm\bar{3}m$, No. 221), with  
experimental lattice parameters of $a=4.035$ \AA~for \ppd\cite{disalvo16,pbpd3}
and $a=3.971$ \AA~ for \spd\cite{snpd3_exp1,snpd3_exp2}.
Both experimental parameters are smaller by about 2\% than our values optimized in GGA.
Here, our calculations were based on the experimental values, unless mentioned otherwise.

The topological properties were explored by the Wannier function approach.
From the band structures obtained from {\sc wien2k}, 
Wannier functions and the corresponding hopping amplitudes were generated 
using the {\sc wannier90}~\cite{marzari} and {\sc wien2wannier}~\cite{jan10} programs.
The surface spectral functions were calculated by the {\sc wanniertools} code.\cite{wtools} 

The transport and optical properties were calculated by two extensions of {\sc wien2k}.
Based on the semiclassical Boltzmann transport theory with a constant scattering time
approximation, the transport calculations were carried out with the {\sc boltztrap} code.\cite{boltz}
Calculation of optical properties including SOC is available in the {\sc optics}.\cite{optic}
Assuming an inverse scattering lifetime $\gamma=10$ meV in the intra-band contribution,
the dielectric function $\epsilon(\omega)$, contributed by both intra- and inter-band
excitations, was calculated. 
The electronic energy loss function is given by --Im$[\epsilon^{-1}(\omega)]$.

In {\sc wien2k}, the BZ was sampled by a very dense $k$-mesh
40$\times$40$\times$40 due to its sharp DOS structure near E$_F$.
For a careful check of the position of the flat band, a high value of basis
set cutoff $R_{mt}K_{max}=9$
was used to determine the basis size with atomic radii of 2.5 bohr 
for the both ions.

\section{Electronic and Topological Characters of  P\lowercase{b}P\lowercase{d}$_3$}

\begin{figure*}[tbp]
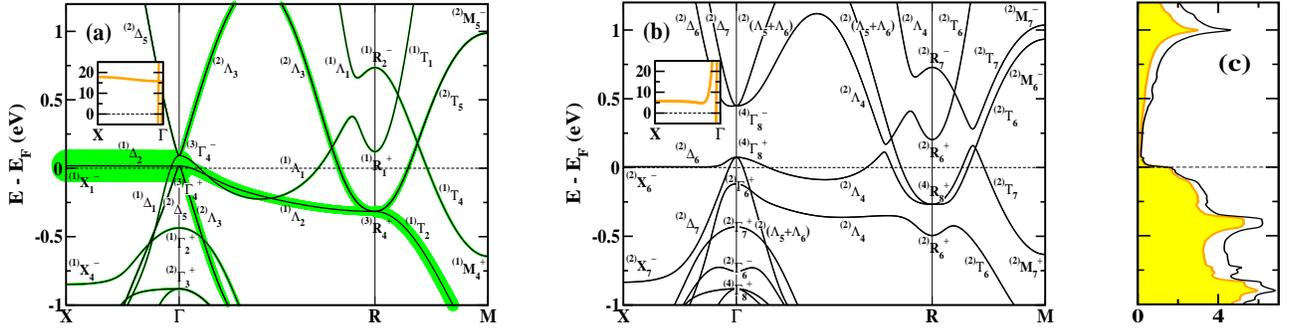

{\resizebox{6.5cm}{4.3cm}{\includegraphics{Fig2a.eps}}}
\hskip 8mm
{\resizebox{6.5cm}{4.3cm}{\includegraphics{Fig2b.eps}}}
\hskip 9mm
{\resizebox{2cm}{4.3cm}{\includegraphics{Fig2c.eps}}}
\caption{Enlarged view of the (a) GGA and (b) GGA+SOC band structures 
labeled with with point group representations,
along the $X-\Gamma-R-M$ line around $E_F$, which is set to zero.
The superscripts of the notations denote degeneracies of each band.
At the high symmetry points, the $\pm$ symbols indicate parities of each band.
The $\Gamma-R$ line has a $C_{3v}$ symmetry, while the other lines 
have $C_{4v}$ symmetry.
The insets show the flat band lying a few meV above $E_F$ along the $\Gamma-X$ line. 
The symmetry points of the band structures are given in Fig.~\ref{str}(b).
The $X$-point is the zone boundary along the (100) direction. 
(c) The GGA+SOC density of states (DOS) of \ppd, in units of states/eV.
The (yellow) shaded region corresponds to the orbital-resolved DOS of Pd $5d$ states.
In (a), the $d_{yz}$ characters of Pd2 ($\frac{1}{2}\frac{1}{2}0$) 
and Pd3 ($\frac{1}{2}0\frac{1}{2}$) ions are highlighted by thick (green) lines.
Note that there is no contribution of Pd1 ($0\frac{1}{2}\frac{1}{2}$) ion to the 
flat band along this direction.
}
\label{pb_band}
\end{figure*}

\begin{figure*}[tbp]
{\resizebox{16cm}{5cm}{\includegraphics{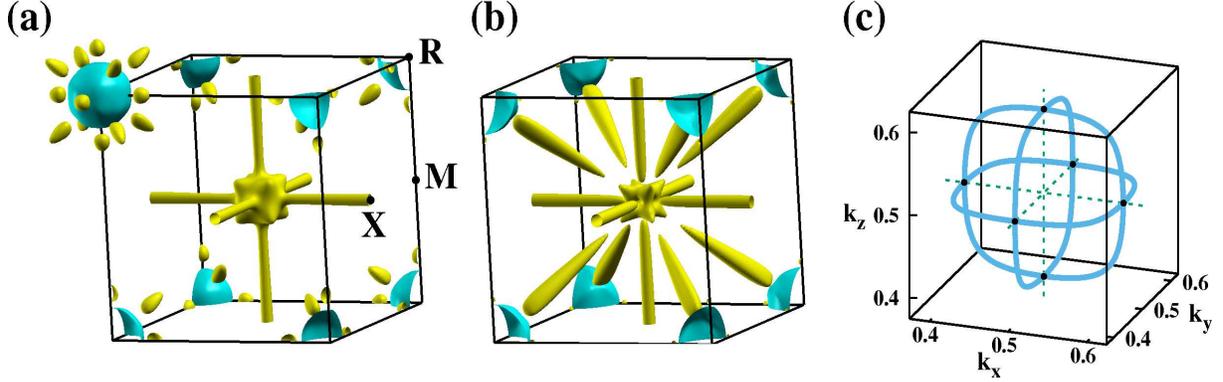}}}
\caption{GGA+SOC Fermi surfaces of (a) \ppd~ and (b) \spd.
For \ppd, the $R$-centered sphere (blue) contains electrons, while the others 
(yellow) enclose holes.
The 14 Dirac points in the \ppd~BZ lies along the $R-\Gamma$ and $R-M$ lines,
coincidentally between the $R$-centered sphere and the carrot-like 
and spheroid hole pockets, see panel (a).
In \spd, large carrots appear along the (111) directions.
(c) Nodal line isocontours connecting the Dirac points, denoted by black dots, at $\sim$0.5 eV
 along the $M-R-M'$ line  in GGA.  The center of the fractional BZ is the $R$-point. 
}
\label{fs}
\end{figure*}

\begin{figure*}[tbp]
{\resizebox{15cm}{10cm}{\includegraphics{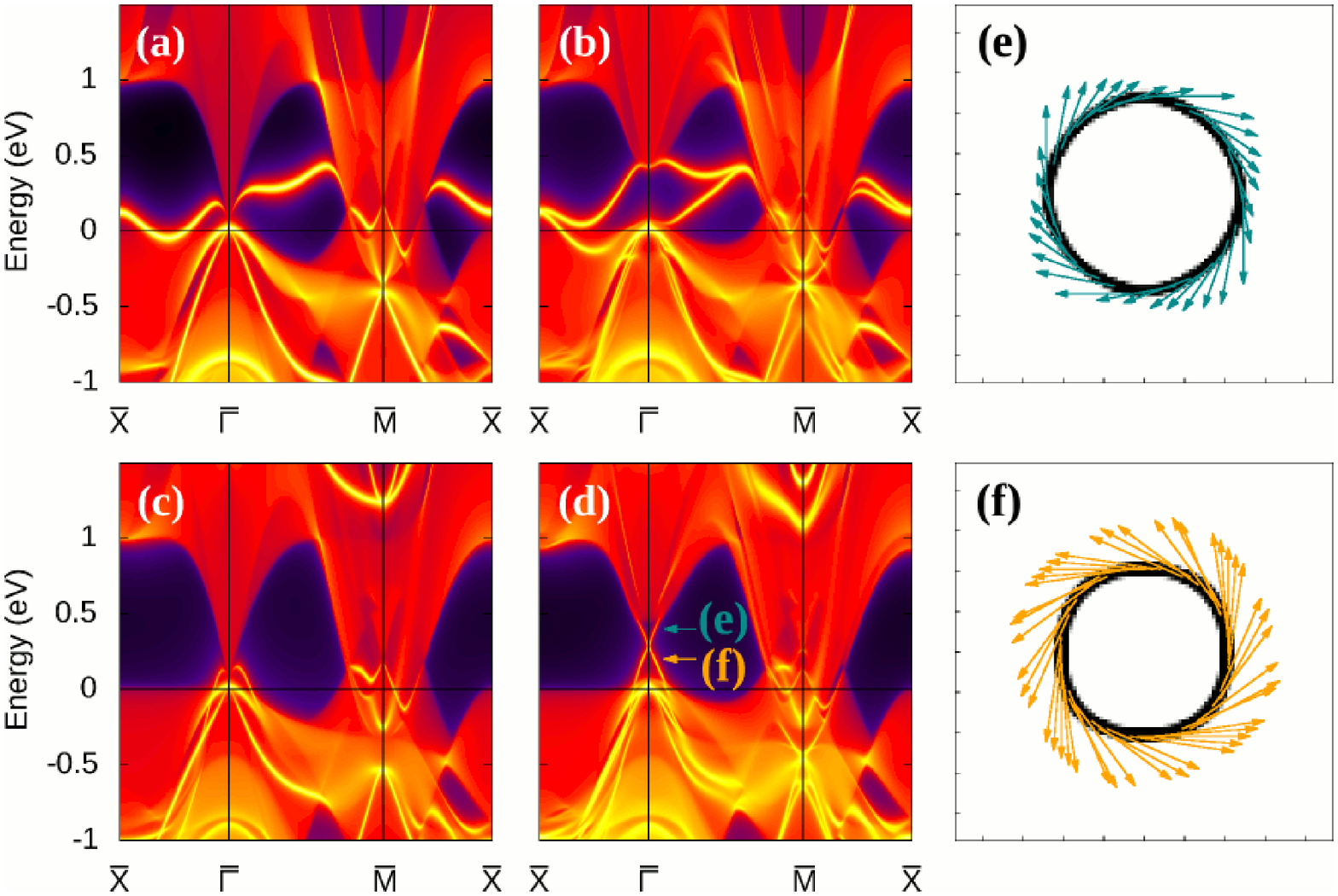}}}
\caption{The (001) surface spectral functions for (top) Pd$_2$ and 
(bottom) Pb-Pd terminations within (left) GGA and (middle) GGA+SOC. 
The reddish regions denote projections of bulk states.
The right panels display the spin texture at the two-dimensional Dirac cone appearing in (d) 
at $\bar{\Gamma}$, having opposite helical texture in the (e) upper and (f) lower cones.
The surface projected symmetry points are given in Fig. \ref{str}(b).
}
\label{sf}
\end{figure*}

\subsection{Electronic structure of PbPd$_3$}
The band structures of both GGA and GGA+SOC near $E_F$ are shown 
in Figs. \ref{pb_band}(a) and \ref{pb_band}(b).
In this energy regime, the bands have mostly Pb $6s$, $6p$ above E$_F$ and Pd $5d$ 
character below.
The corresponding density of states (DOS) with Pd $5d$ orbital-projected DOS 
is given in Fig. \ref{pb_band} (c).

The band structure shows unusual characteristics relating to transport, thermodynamic, and topological
properties. Most prominently, a very flat band of Pd $5d$ character appears along 
the $\Gamma-X$ line only a few meV above $E_F$.
The lack of dispersion reflects the lack of nearest neighbor (NN) Pd $dd\delta$ hopping.
This characteristic appears commonly in cubic perovskite-related systems,\cite{LP09}
but is unusual in intermetallic compounds.
As indicated in Fig.~\ref{pb_band}(a), for the $\Gamma-X$ line along the (100) direction, 
the flat band has solely the $d_{yz}$ orbital character
of  Pd2 ($\frac{1}{2}\frac{1}{2}0$) and Pd3 ($\frac{1}{2}0\frac{1}{2}$) ions.
Significantly, in \ppd~ this flat band lies only 6 meV above $E_F$, thereby giving an
effective Fermi energy for holes of E$_{F,h}^{*}$= 6 meV and corresponding Fermi temperature
T$_{F,h}^{*}$=70 K.
As can be seen in Fig. \ref{pb_band}(c), 
structure in the DOS is a 2D-like step discontinuity, and electron doping of only $\sim$0.01
carrier/f.u. would be required to move E$_F$ up to the band edge. 
Above the edge, the DOS is very small (semimetal-like) up to 0.2 eV.

Another aspect that we will follow is an occurrence of 3-fold degeneracies along
symmetry lines, which are becoming known as triple nodal points (TNPs).
In the absence of SOC, as shown in Fig. \ref{pb_band}(a), 
spinless TNPs $\Gamma_4^\pm$ appear at the $\Gamma$ point.
Along the $\Gamma-R$ line, having the $C_{3v}$ symmetry,
the $\Gamma_4^-$ band at 0.1 eV splits into a doublet $\Lambda_3$ of a positive effective mass 
and a singlet $\Lambda_1$ of a negative effective mass.
The $\Gamma_4^+$ band slightly above $E_F$ splits into a doublet $\Lambda_3$
and a singlet $\Lambda_2$, both with negative effective masses.
As previously noted,\cite{brad16,zhu16} 
the $\Lambda_3$ and $\Lambda_1$ crossing forms a TNP at 
(0.39,0.39,0.39)$\frac{2\pi}{a}$ at 0.15 eV
along the $\Gamma-R$.

At the $R$-point, a TNP of $R_4^+$ symmetry lies at --0.32 eV.
This band splits into a doublet $T_5$ of a positive effective mass and
a singlet $T_2$ along the $R-M$ line, having $C_{4v}$ symmetry.
This doublet forms a TNP with the $T_4$ at 
$(\frac{1}{2},\frac{1}{2},0.312)\frac{2\pi}{a}$ at 0.19 eV. 
The TNP along the $C_{4v}$ symmetry line has not been discussed previously.
One may expect surface states connecting these spinless TNPs in the absence of SOC.
The surfacce states for \ppd~ are discussed below. 

In addition to TNPs, there is a nodal point at $\sim 0.5$ eV along the $R-M$ line at 
$(\frac{1}{2},\frac{1}{2},0.3983)\frac{2\pi}{a}$.
Perpendicular to this line and along cubic directions, this double 
degeneracy {\it does not split}, forming 
the three interconnected topological nodal loops shown in Fig. \ref{fs}(c).
The three nodal loops lie on mutually perpendicular mirror planes.
These $R$-centered links are similar to what was reported in the 
antiperovskite Cu$_3$PdN.\cite{cu3pdn}  

To probe the origin and sensitivity to symmetry lowering of these TNPs, 
the cubic structure has been distorted, conserving the volume.
Tetragonal distortion leads to breaking $C_{3v}$ symmetry, thereby 
removing TNPs along the (111) direction.
Both the $C_{3v}$ and $C_{4v}$ symmetries are broken by orthorhombic distortion,
resulting in the vanishing of the TNPs along both the $R-M$ line as well as 
along the $\langle 100\rangle$ directions. 
See the Supplemental Material for additional information.\cite{suppl}
These TNPs are protected by the mirror plus three-fold, 
or four-fold, rotational symmetries respectively.

Comparison of the two band structures without and with SOC in Fig. \ref{pb_band}
reveals several effects of SOC.
SOC leads to anticrossing of several, but not all, crossing bands,
as well as lowering the flat band toward $E_F$, from 15 meV to 5 meV.
At the $\Gamma$ point, two 6-fold degenerate bands $\Gamma_4^\pm$  (including spin) split into
doublet and quartet, resulting in a SOC gap of 0.3 eV due to a band inversion.
Substantial impacts of SOC resulting in changes of surface states
appear in the spinless TNPs along $\Gamma-R-M$ lines
near the $R$-point at $\sim$0.2 eV.
As discussed by Zhu and coworkers,\cite{zhu16}
these TNPs are not protected in the cubic structure, when including SOC.
Instead, SOC driven anticrossings at these TNPs lead to eight symmetry related
fourfold degenerate Dirac points
at $\sim$0.2 eV along $\Gamma-R$,  and six at $\sim$0.15 eV along $R-M$.
A similar behavior along a $C_{6v}$ line was suggested 
in the layered hexagonal Al$B_2$-type metal diborides.\cite{zhang17} 
Thus the band structure is gapped at 0.2 eV except for the band
emanating from $R$, strengthening the semimetal
viewpoint of the near-E$_F$ electronic structure of \ppd.

One result of this semimetallic band structure is delicate Fermi surfaces (FSs).  
Figure \ref{fs} displays the physical FSs (SOC included), 
consisting of three types. Narrow cylindrical open hole FSs lie along the (100) axes, 
intersecting at the zone center;  these tubes are related to the flat band along 
$\Gamma-X$ discussed above.
An $R$-centered electron spheroid FS has a radius of $0.26(\frac{\pi}{a})$,
containing 0.03 electrons per formula unit.
Additionally, there are tiny elliptical hole pockets (``carrots'') around this sphere
along the \{111\} directions and tiny spheroids along the \{100\} directions.

Most results obtained for our optimized lattice parameter
are very close to those for the experiment parameter presented above, 
as expected from the fact the that the difference is small (less than 2\%). 
There is however a distinction worth mentioning.
For the optimized (larger) volume, the flat band is shifted to --5 meV
relative to $E_F$. Thus modest pressure can be used to tune this band edge through the
chemical potential for the stoichiometric composition.
The position of the flat band can also be tuned by either 
tetragonal or orthorhombic strain, with more information provided in
the Supplemental Material.\cite{suppl}

\begin{figure}[tbp]
{\resizebox{7cm}{7.5cm}{\includegraphics{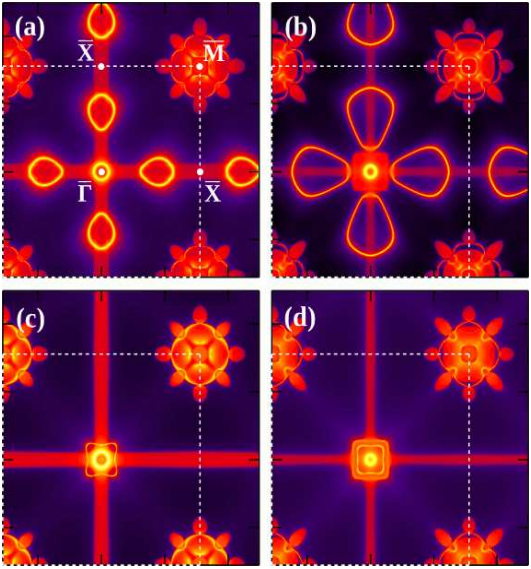}}}
\caption{Spectral densities of the surface Fermi contours of the 
surface states shown in Fig. \ref{sf},
for (top) Pd$_2$ and (bottom) Pb-Pd terminations. 
The left panels are for GGA only, the right panels show results for GGA+SOC.
The strong yellowish lines denote surface bands.
}
\label{sf_fs}
\end{figure}

\begin{figure}[tbp]
\vskip 8mm
\resizebox{7.5cm}{6cm}{\includegraphics{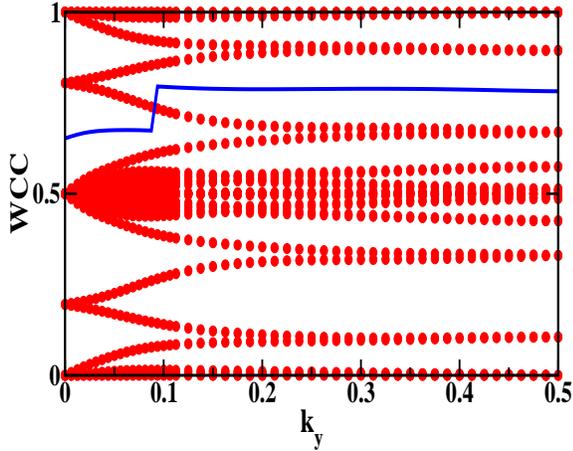}}
\caption{Plot of the hybrid Wannier charge centers (WCCs) plot (red, thick lines)
 across half of the Brillouin zone in the $k_z=0$ plane, 
 showing an odd number of crossings between the charge center 
 and largest gap among two adjacent WCCs.
 The gap function is designated by the blue (thin) line.
 The magnitude of the wave vector $k_y$ in the horizontal axis is given in units of $\pi/a$.
}
\label{wcc}
\end{figure}

\subsection{Topological character of PbPd$_3$}
Using a Wannier representation of the bands, 
we have carried out surface calculations based on the Green function approach\cite{green}
for a semi-infinite crystal.
For \ppd, there are two possible (001) surface terminations, one containing both Pb
and Pd atoms (Pb-Pd) and the other containing only Pd surface atoms (Pd$_2$).
Figure \ref{sf} shows the surface spectral functions for the each termination, 
in both GGA and GGA+SOC.
Around 0.2 eV above E$_F$ the surface zone is gapped except around the $\bar{M}$ point,
arising from the bulk bands around $R$ that project onto $\bar{M}$. 

In the absence of SOC, in Figs. \ref{sf}(a) and (c)
the Pd$_2$ termination shows a weakly dispersive surface band in both directions from
$\bar \Gamma$. 
For the Pb-Pd terminated surface only surface resonances within the bulk bands
appear along symmetry lines. 
Several surface states appear at the projections of TNPs
both in the gap and inside the bulk states in both terminations.
The positions of TNPs and their projections onto the surface BZ are 
provided in Fig. \ref{str}(b).
At $\bar{\Gamma}$ there is a surface band $\sim$ 15 meV above $E_F$, corresponding
to the TNP at $\Gamma$ at 16 meV that projects onto $\bar{\Gamma}$.
One of the TNPs (at 0.2 eV) along the $R-M$ line projects onto $\bar{M}$, and any
surface states that might be related are hardly evident, as it is within a strong
background of projected bulk states around $\bar{M}$.
The other two TNPs project onto the $\bar{M}-\bar{X}$ line, where a strong surface band
appears within the gap at 0.1-0.3 eV, but only for the Pd$_2$ termination. This strong
surface band is seen everywhere within the gap. Since it is missing for the PbPd
terminated surface, its relation to TNPs is in doubt.  

SOC is known to have strong consequences in nodal point and nodal loop materials.
Inclusion of SOC leads to a large 0.4 eV gap just above E$_F$ at the $\Gamma$ point
and all along the $\Gamma-X$ line, hence a 0.4 eV gap at $\bar{\Gamma}$,
resulting in qualitative changes in the surface spectrum.  As shown in Fig. \ref{sf}(b), 
in the Pd$_2$ termination the surface band in the absence of SOC becomes two surface
bands, separated by a 0.4 eV gap around $\bar{\Gamma}$ point but connecting
valence and conduction bands.
These bands merge at $\bar{X}$ and also as they enter the bulk spectrum along the
other two symmetry lines.

For Pb-Pd termination, where no surface bands around $\bar{\Gamma}$ appear within the gap in GGA, 
SOC produces a new occurrence, one that is very different from that of the Pd-Pd 
termination: a surface band Dirac point emerges at $\bar \Gamma$ 
at 0.25 eV above E$_F$ from two linear bands joining valence to conduction
bands, as shown in Fig. \ref{sf}(d). 
This surface crossing is enabled by SOC opening a bulk gap at $\Gamma$.
In the Pb-Pd termination, a 2D Dirac cone at the $\bar{\Gamma}$ point 
is a dominant feature.
In Fig. \ref{sf} (e,f) the spin textures are shown for cross sections of the
cones just above, and just below, the Dirac point.  The orientation of texture
is in opposite helical direction
in the upper and lower cones, reflecting the topological character of this point.

\subsection{Fermi level spectral density}
The surface Fermi level spectral density contours (the surface Fermi lines)
are pictured in Fig. \ref{sf_fs}, presented analogously to Fig. 4.
As expected from differences between the two terminations in the surface spectrum,
a clear distinction appears along the $\bar{\Gamma}-\bar{X}$ line,
whereas features around $\bar{M}$ are nearly identical.
For Pd$_2$ termination there are four symmetry related ovoid pockets along the
axes, both without and with SOC. These are shown in Fig. \ref{sf_fs} (a) and (b),
indicating that SOC substantially increases the area of these pockets (increases
the density of the corresponding carriers).
The bulk $R$ point and its Fermi spheroid project onto the $\bar{M}$ point,
and protrusions along axes and diagonals connected to this spheroid appear for 
both terminations. There is only small quantitative change with SOC.
For higher resolution figures, see the Supplemental Material.\cite{suppl}

Calculations of the hybrid Wannier charge centers (WCCs) allow a ``theoretical spectroscopy''
of topological character in cases where the full zone is not gapped.\cite{hwcc,z2pack} 
In this technique, the Bloch-to-Wannier transformation is 
applied only to one direction, and in cubic crystals there is only one choice of direction.
These calculations were performed to establish topological character 
using the {\sc wanniertools} code.\cite{wtools}
In this method, the $Z_2$ number is calculated from
the number of crossings of WCCs mod 2 by an arbitrary line in half of BZ,
making use of time-reversal symmetry.\cite{hwcc}
As a practical matter, instead of an arbitrary line, a line between two adjacent WCCs
is chosen for numerical efficiency.

In \ppd, inclusion of SOC separates bands over most of the BZ, 
band crossings surviving only around the $R$ point.
Thus the electronic structure on the basal planes (viz. $k_z=0$) 
can be considered as insulating subspaces, thus providing a well-defined $Z_2$ number
that can be calculated using the WCC method.
As shown in Fig. \ref{wcc}, there is an odd number of crossings of the largest gap
with the charge centers,
giving $Z_2$=1.
This number is also obtained from parities of all occupied eigenstates at time-reversal
invariant momenta in the $k_z=0$ plane. 
The parities are given by --1 at $\Gamma$ and +1 at $X$ and $M$,
for a product of --1, again indicating $Z_2$=1.
Therefore this crystal is a topological insulator on the $k_z=0$ plane.

\subsection{Strain effects}
We  have considered tetragonal and orthorhombic distortions to investigate effects of strain.
The $Z_2$ indices are well defined in both $k_z$=0 and $\pi/c$ planes in both cases
even for a small strain. 
Due primarily to a change of parity at $\Gamma$, 
the corresponding indices are 0;(0,0,0) for the both cases, 
indicating a topologically trivial phase. 
However, in-gap surface states appear,
similar to the topological crystalline insulating phase
proposed in the tetragonal lattice\cite{fu11} or the orthorhombic perovskite 
iridates.\cite{kee15}
For more information see the Supplemental Material.\cite{suppl}
 
Two sister compounds, semimetallic SnPt$_3$ and PbPt$_3$, are worthy of comment. 
In contrast to the Pd analogies with several TNPs above $E_F$,
the valence bands are fully filled, so SOC leads to well separated bands around $E_F$
throughout the BZ.\cite{pbpt3_our}
SnPt$_3$ has been proposed as a weak topological insulator,\cite{snpt3_ti}
while PbPt$_3$, with its additional complexities,  will be discussed 
elsewhere.\cite{pbpt3_our}

\begin{figure}[tbp]
{\resizebox{8cm}{5cm}{\includegraphics{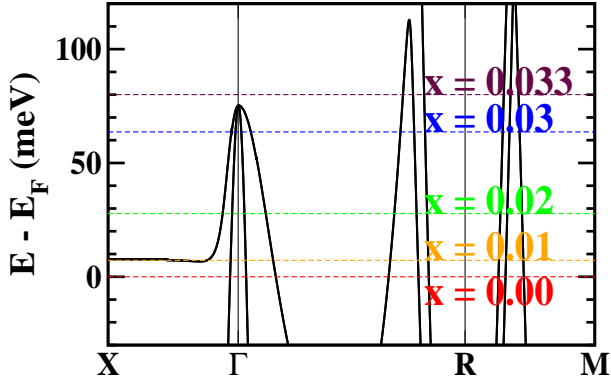}}}
\caption{Enlargement of the band structure in GGA+SOC, with horizontal
lines denoting $E_F$ as electron doping $x$ is increased. The results are
from the virtual crystal approximation applied to (Pb$_{1-x}$Bi$_x$)Pd$_3$.
1\% doping puts $E_F$ precisely on the flat band.
}
\label{doped}
\end{figure}

\section{Role of the flat band; electron doping}
We have varied the electron concentration in (Pb$_{1-x}$Bi$_x$)Pd$_3$, 
which is experimentally accessible,\cite{pbpd3} 
to inspect the role of the flat band in transport and optical properties, 
using the virtual crystal approximation, for which this is an optimal application.
We will focus on GGA+SOC results.
There is negligible change in band structure, and Fig.~\ref{doped} displays the 
rising of the Fermi level with added electrons.
The flat band precisely lies at $E_F$ for $x=0.01$,
and by $x=0.02$ the intersecting cylinder FSs disappear.
At $x=0.033$, the hole pockets at the $\Gamma$ point are filled.
Since transport and optical properties are derived from the bands
$\varepsilon_k$, 
its derivatives, and the Fermi level position,\cite{boltztheory}
one may anticipate specific features related with the flat band in \ppd.
These properties are surveyed in the following subsections.

\begin{figure}[tbp]
{\resizebox{8cm}{8cm}{\includegraphics{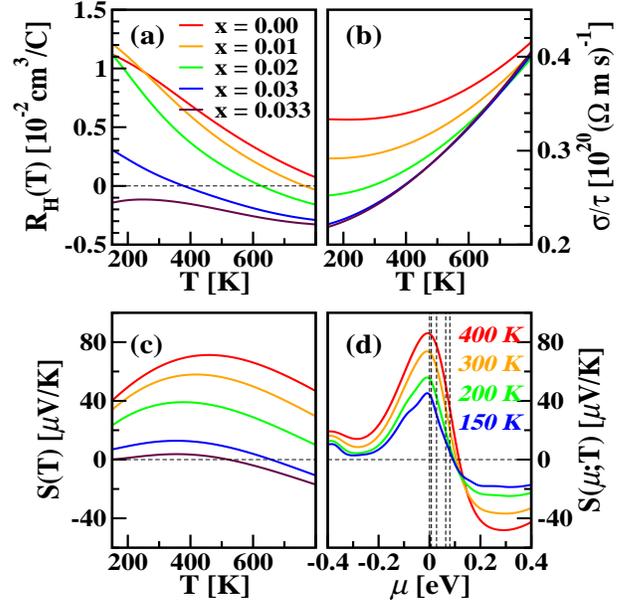}}}
\caption{As varying doped-electron concentration,
 temperature-dependent (a) Hall coefficients R$_H$(T),
 (b) electronic conductivity $\sigma$ over scattering time $\tau$,
 and (c) Seebeck coefficients $S$ for GGA+SOC in the range of 150 K -- 800 K.
 (d) Chemical potential $\mu$-dependent Seebeck coefficients $S(\mu;T)$ 
 for various temperatures in GGA+SOC.
 In (d), the vertical dashed lines indicate $\mu$ at each concentration,
 in the order of $x=0$ to $0.033$ from the left.
}
\label{boltz}
\end{figure}

\subsection{Transport properties}
The transport properties have been calculated
based on quasiclassical Bloch-Boltzmann transport theory,\cite{boltztheory} assuming a 
constant scattering time $\tau$ approximation. Our results including SOC
are pictured in Fig.~\ref{boltz}.
The Hall coefficient $R_H(T)$, given in Fig.~\ref{boltz}(a), 
shows substantial variation with temperature even at these very low doping levels.
At zero doping $x$=0, it is net (from competing Fermi surfaces) electron-like in sign.
With increasing $T$, $R_H(T)$ monotonically decreases, 
and just above 800 K crosses zero, where electron and hole contributions compensate.
With increasing doping, $R_H(T)$ crosses zero at lower temperatures, {\it e.g.},
for $x=0.03$  exact compensation of holes and electrons occurs at 400 K.
For $x=0.033$, $R_H(T)$ is negative for the whole range of T, 
indicating dominance of hole carriers.
Recall however that in a multiband system $R_H(T)$ has no simple relation
to carrier densities.\cite{nbp}

Figure \ref{boltz}(b) shows the $T$-dependent conductivity over scattering time $\sigma/\tau$.
For a typical metal with large N(E$_F$), $\sigma/\tau$ is proportional to 
$N(\mu)\langle v_F^2\rangle$\cite{singh10}
with insignificant $T$-dependence. 
Here $N(\mu)$ and $\langle v_F^2\rangle$ are the DOS at the chemical potential $\mu(T)$
and the thermal average of square
of the Fermi velocity at $\mu(T)$, respectively. With structure in $N(E)$ around E$_F$,
$\mu$ becomes T-dependent.
As T rises above 400 K ($x$=0),  $\sigma/\tau$ increases, by about 25\% at 800K,
due to the sharp drop in $N(E)$ at 6 meV above E$_F$. As doping increases, the low temperature
value drops and the temperature increase is enhanced as E$_F$ moves upward with doping, 
through the DOS singularity and into the semimetal region.

Related variations are reflected  
in the Seebeck coefficient $S(T,x)$,\cite{singh10}  given by
\begin{eqnarray}
S(T,x)=\frac{\pi^2}{3}\frac{k_B^2 T}{e}\left(\frac{d\mathtt{ln} 
          \sigma(\varepsilon;T,x)}
                                         {d\varepsilon}\right)_\mu .
\label{eqn1}
\end{eqnarray}
As $T$ increases, $S(T,x)$ increases up to about 300-500K depending on $x$ and then drops,
as displayed in Fig. \ref{boltz}(c). For $x$=0.033, $S(T,x)$ is very small, a condition that
is poor for thermoelectric applications but
can be favorable for other applications.
Fig. \ref{boltz}(d) presents a different viewpoint, plotting $S(\mu,T)$ versus $\mu$ at
four values of T. At each temperature, the maximum occurs at $\mu(T)$=0, followed by a
steep drop through zero around $\mu$=0.1 eV, before leveling off beyond $\mu$=0.2 eV. 
Measurement of the Seebeck coefficient, which is independent of scattering in the constant
scattering time approximation, could be useful in determining the stoichiometry of
\ppd~ at the 1\% level.

Although interesting features are visible in the $x$, $\mu$, and T dependences of the
transport coefficients, there is nothing that is as striking as might have been expected
given the anomalous $N(E)$ very near E$_F$. 
A contributing feature is that 
the FSs are small and varied in size and shape, and the Fermi velocities are small:
on the crossed cylinders  $v_F\sim 6.4\times10^6$ cm/sec, compared to values 20 times 
larger for many intermetallic compounds.

\begin{figure}[tbp]
{\resizebox{7cm}{5cm}{\includegraphics{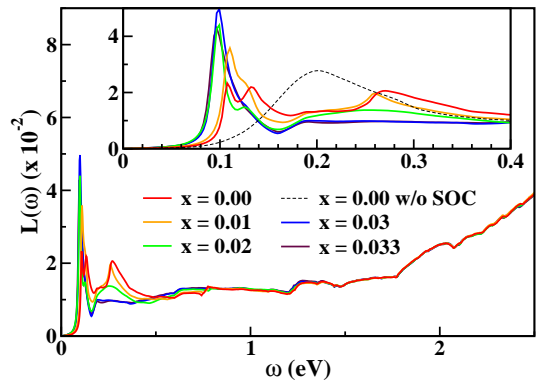}}}
\caption{Energy loss function $L(\omega)$ versus energy $\omega$ ($\hbar$=1),
for the various electron-doping levels.
A scattering rate $\gamma=\hbar/\tau$ of 10 meV has been assumed, which primarily affects
the widths of peaks but also determines the (small) intraband contribution.
Inset: Enlarged plot below $\omega=0.4$, showing the low energy plasmonic peak(s) clearly.
For $x$=0, the result without SOC is included to indicate the influence
of SOC in the loss function.
}
\label{optic}
\end{figure}

\subsection{Optical properties}
The electron energy loss function $L(\omega)$, given in Fig. \ref{optic}, 
was calculated as the imaginary part of the inverse dielectric function $\epsilon(\omega)$.
The dielectric function includes both intra- and interband contributions.
In the intraband part, containing the Drude term, 
an inverse scattering time $\tau$ is chosen from $\gamma=\hbar/\tau$=10 meV.
 
The overall behavior of $L(\omega)$ is non-monotonically dependent on electron-doping level.
\ppd~ is a nearly transparent (semi)metal in the far infrared region $\omega<75$ meV.
At zero doping, the large loss peak at 0.2 eV when SOC is neglected is split by SOC, 
leaving a series of less pronounced subpeaks
at 0.11, 0.14, and 0.27 eV.
The plasmon peak around 0.1 eV is mainly due to excitation at the $\Gamma$-point
and has the strongest intensity at $x=0.03$.
The intensity and position of these peaks depends somewhat on 
the choice of $\gamma$.

\begin{figure}[tbp]
{\resizebox{8cm}{4cm}{\includegraphics{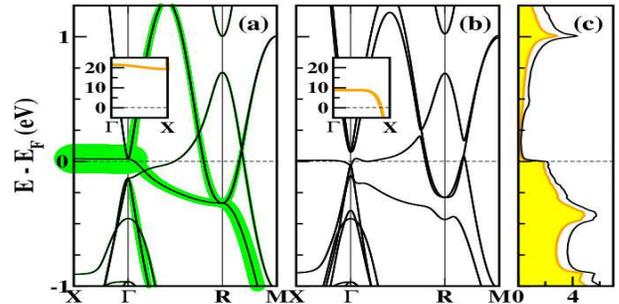}}}
\caption{The band structure and density of states of \spd, as 
labeled, which can be compared with that of PbPd$_3$ in Fig. \ref{pb_band}. 
}
\label{sn_band}
\end{figure}

\section{The S\lowercase{n} analog}
Isovalent and isostructural \spd, with a 1.5\% smaller lattice constant than
\ppd, was also investigated.
As expected, the band structure of \spd, presented in Fig. \ref{sn_band}(a),
is very similar to that of \ppd, using GGA only.
However, in \spd~ effects of SOC are not as substantial as in \ppd, 
since the strength of SOC in Sn is weaker than in Pb.
The GGA+SOC band structure and the corresponding DOS are displayed 
in Figs. \ref{sn_band}(b) and (c).
Compared to \ppd, there are two distinctions worth noting. First,
the flat band, which is shifted downward by SOC, lies at 10 meV versus 6 meV for \ppd.
Second, the bottom conduction band lies very close to, but above, E$_F$ along the 
$\Gamma-R$. 
These two features result in change in fermiology of \spd, shown in Fig. \ref{fs}(b)
beside that of \ppd.
The radius of the pipe-like FS is increased 
and irregular rod-shape FSs appears along the \{111\} directions.

\section{Summary}
Using various calculations based on {\it ab initio} approach and semiclassical
Bloch-Boltzmann theory,
we have investigated the peculiar electronic structure, topological characters,
and some transport characteristics of intermetallic \ppd,
which has a very flat band along the $\Gamma-X$ line nearly coinciding with
the Fermi level.
In the absence of SOC, several triple points emerge near $E_F$ 
along the $C_{3v}$ and $C_{4v}$ symmetry lines, leading to surface Dirac cones
along the $\bar{\Gamma}-\bar{M}$ lines and Fermi arcs around the zone corner 
on the surface states.
Nodal points along the $M-R-M'$ line lead to three dimensional nodal links 
that lie on mirror planes.
SOC removes degeneracies in most of the zone, 
leading to a topological insulating phase on the $k_z=0$ plane.
We have furthermore inspected effects of the unusual dispersionless band 
on transport and optical properties by varying the doping level. 

Although there are some distinctions due to difference in 
strength of SOC between Pb and Sn ions, 
most of the electronic structure characteristics of \ppd, including the
flat band, are shared with isovalent and isostructural \spd,
suggesting it will display similar transport, optical, and topological properties.

\section{Acknowledgments}
We acknowledge T. Siegrist for discussions of the phase diagram of \ppd, 
and Y.-K. Kim for useful discussion on the topological properties.
This research was supported by NRF of Korea Grant No. NRF-2016R1A2B4009579
(K.H.A and K.W.L), and by DOE grant DE-FG02-04ER46111 (W.E.P)

\section{Supplemental Material}

Additional details that were mentioned in the main text are provided
here, with description in the figure captions.

\begin{figure*}[htbp]
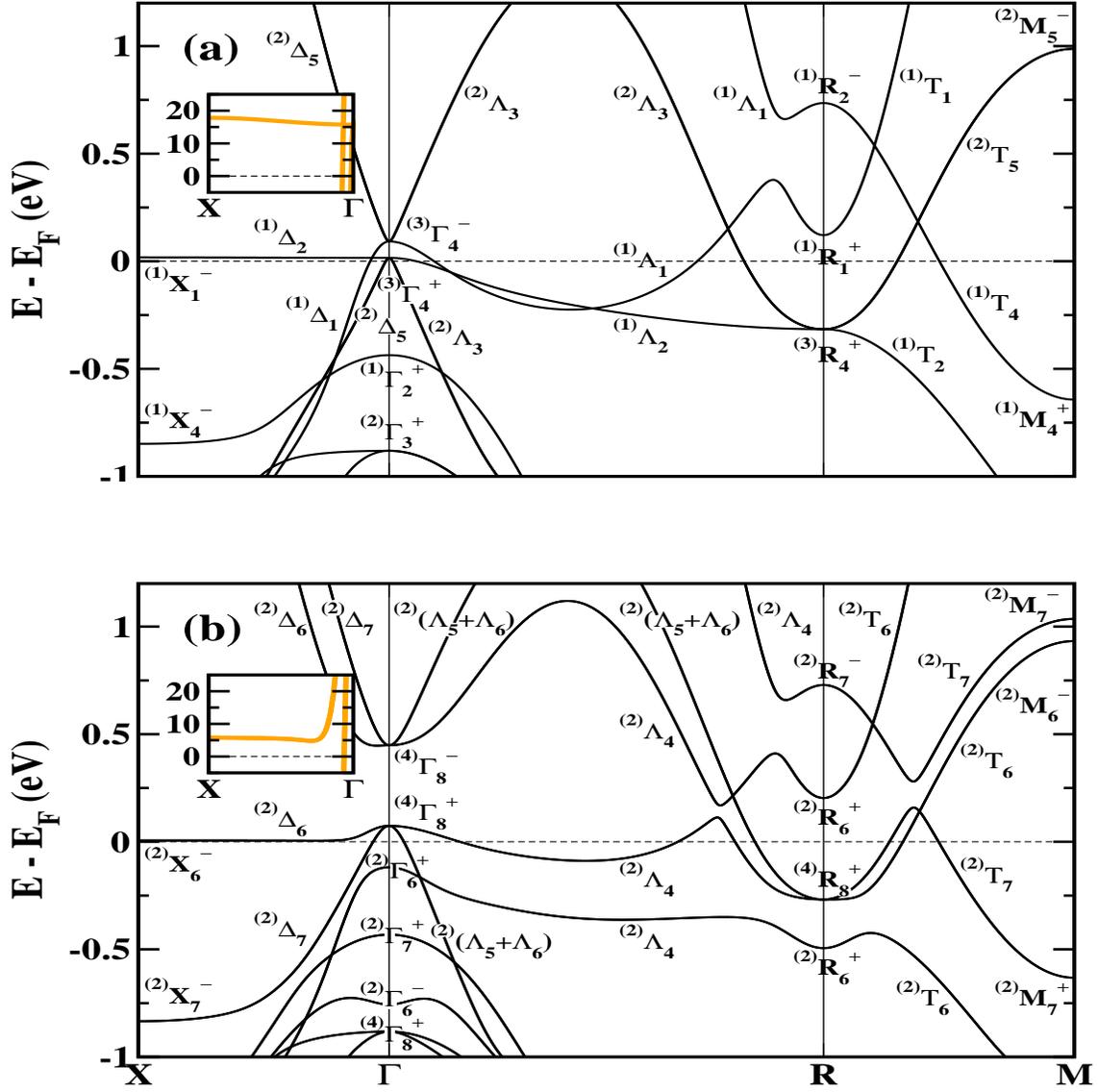

\vskip 10mm
{\resizebox{15cm}{7cm}{\includegraphics{FigS1a.eps}}}
\vskip 10mm
{\resizebox{15cm}{7cm}{\includegraphics{FigS1b.eps}}}
\caption{Blowup plots corresponding to Figs. 2a and 2b in the main text.
}
\label{ns1}
\end{figure*}

\begin{figure*}[htbp]
{\resizebox{12cm}{8cm}{\includegraphics{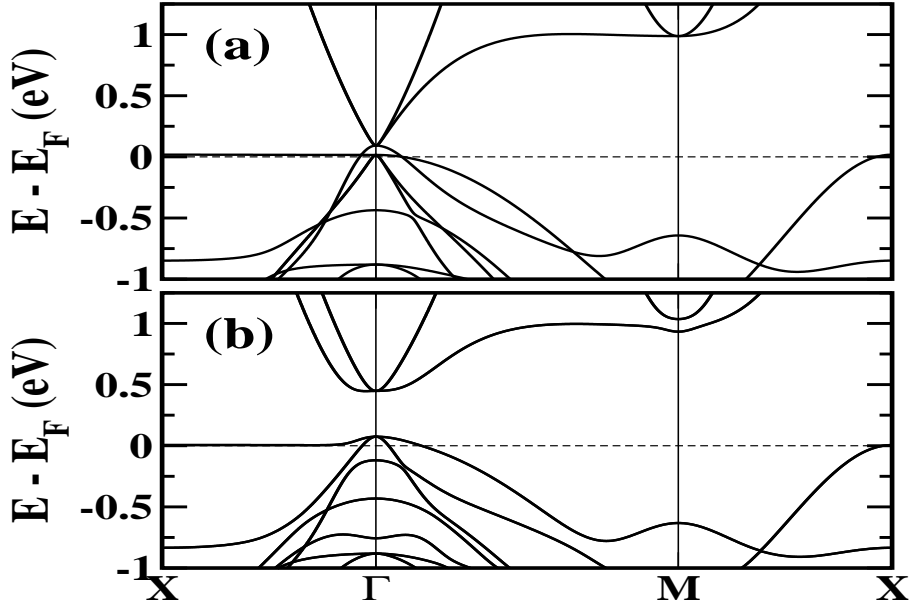}}}
\caption{Band structures, corresponding to Figs. 2a and 2b, 
 along the $X-\Gamma-M-X$ line.
}
\label{ns2}
\end{figure*}

\begin{figure*}[htbp]
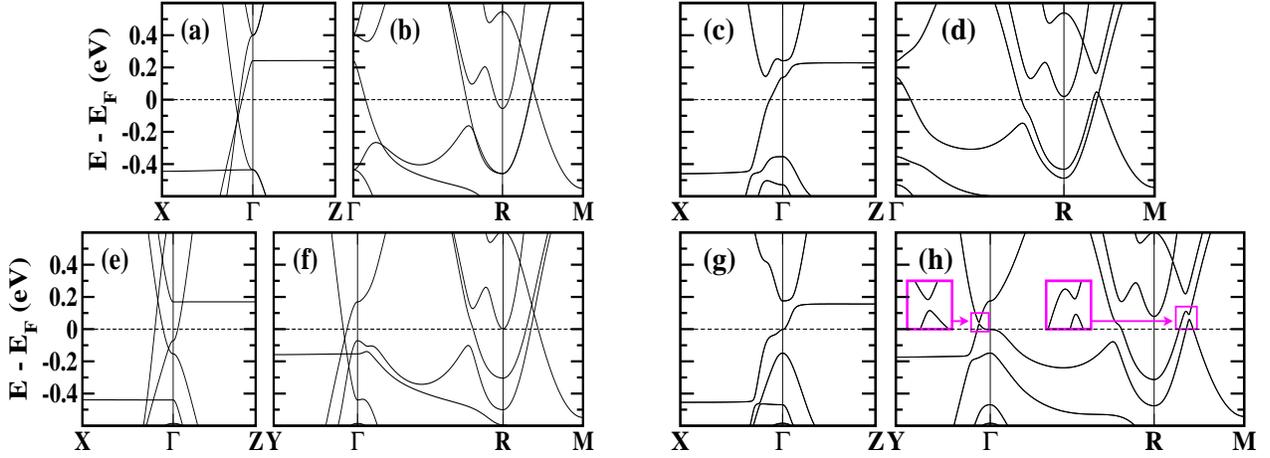

\vskip 8mm
{\resizebox{7.8cm}{6cm}{\includegraphics{FigS3a.eps}}}
\hskip 9mm
{\resizebox{7.8cm}{6cm}{\includegraphics{FigS3b.eps}}}
\caption{Enlarged band structures of PbPd$_3$ for tetragonally (top) 
and orthorhombically (bottom) distorted structures.
The panels of (a), (b), (e), and (f) are in GGA, 
while the panels of (c), (d), (g), and (h) in GGA+SOC.
For comparison, the notation of the high symmetry points follows that of the cubic case.
The $X$, $Y$, and $Z$ points are the zone boundary of (100), (010), and (001), respectively.
For the tetragonal case, the $C_{3v}$ symmetry is broken along (111) direction.
In the orthorhombic case, the $C_{4v}$ symmtery is also broken along (110) direction
at $k_z=\pi/c$ plane. 
Although anticrossings at TNPs occur evern for a small distortion, 
for a better visualization these figures are given for large distortions:
the ratio of lattice parameters of 1:1.1 for the thetraonal structure 
and of 1:1.05:1.1 for the orthrombic structure.
}
\label{s1}
\end{figure*}

\begin{figure*}[htbp]
{\resizebox{14cm}{7cm}{\includegraphics{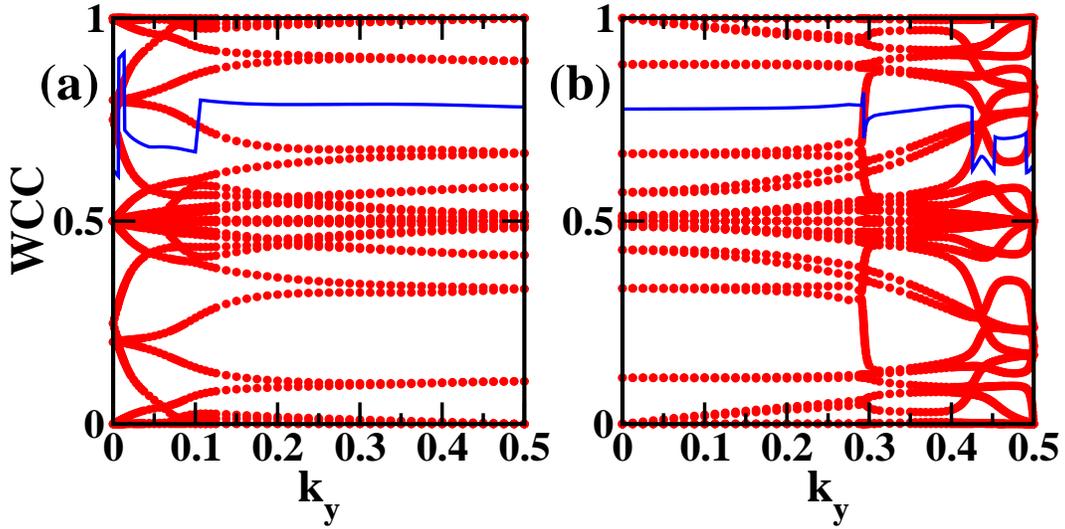}}}
\caption{For the tetragonally distorted structure, 
 plot of the hybrid Wannier charge centers (HWCCs) plot (red, thick lines)
 across half of the Brillouin zone in the (a) $k_z=0$ and (b) $\pi/c$ plane, 
 showing an even number of crossings between the charge center 
 and largest gap (blue, thin line) among two adjacent HWCCs.
The orthorhombic case shows similar behavior, so the figure is not repeated here.
 The magnitude of the wave vector $k_y$ in the horizontal axis is given in the unit of $\pi/a$.
}
\label{s2}
\end{figure*}

\begin{figure*}[htbp]
{\resizebox{14cm}{5cm}{\includegraphics{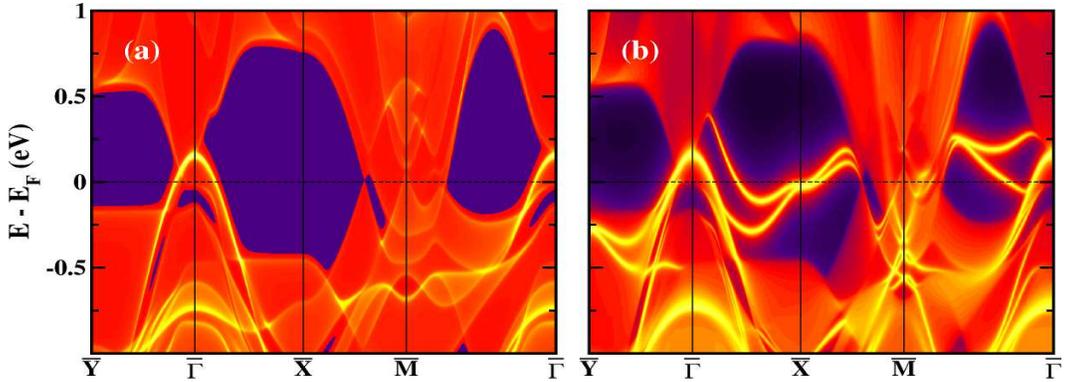}}}
\caption{The bulk-only contribution (left) to the (001) surface spectral function (right) 
 for Pd$_2$ termination of the orthorhombic case in GGA+SOC.
 Compared with two plots, two surface states appear in-gap,
indicating a topological nontrivial state.
}
\label{s5}
\end{figure*}

\begin{figure*}[htbp]
{\resizebox{14cm}{9cm}{\includegraphics{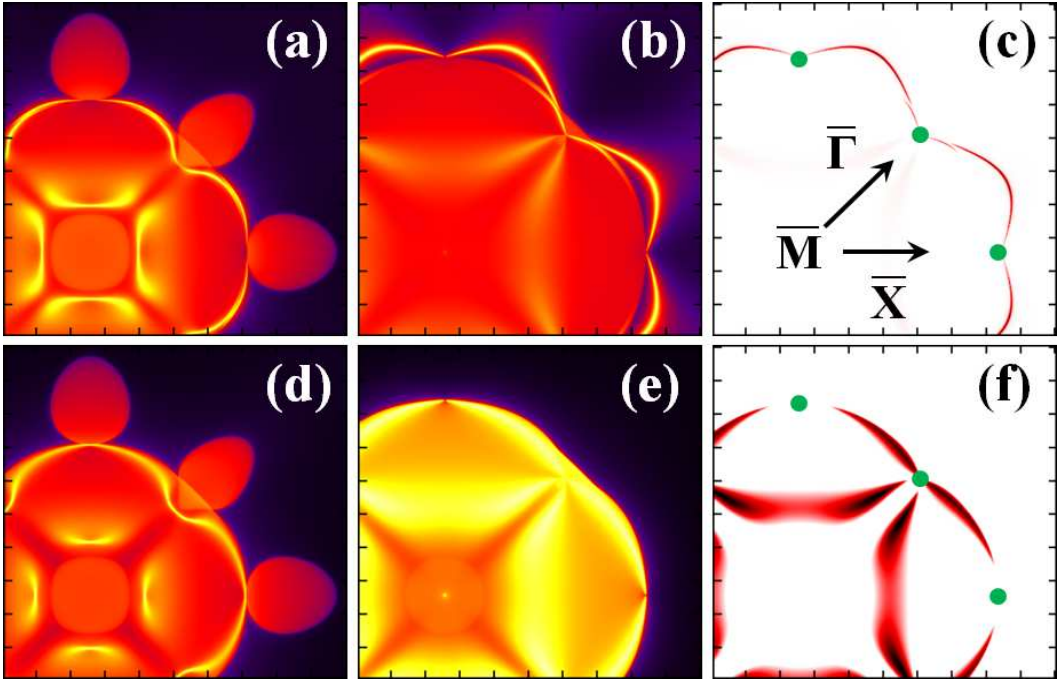}}}
\caption{Enlarged $\bar{M}$-centered (001) surface spectral 
functions for (top) Pd$_2$ and 
(bottom) Pb-Pd terminations in GGA.
The strong yellowish lines denote surface states.
Panels (a) and (d) indicate states at the Fermi energy $E_F$;  (b) and (e) 
indicate states at 0.18 eV where the triple nodal points (TNPs) appear.
Panels (c) and (f) selects only the surface contribution of (b) and (e)
respectively.
The R-centered spheroid is connected to four protrusions by Fermi arcs.
In panels (c) and (f), the green dots denote TNPs.
}
\label{s3}
\end{figure*}

\begin{figure*}[htbp]
{\resizebox{14cm}{9cm}{\includegraphics{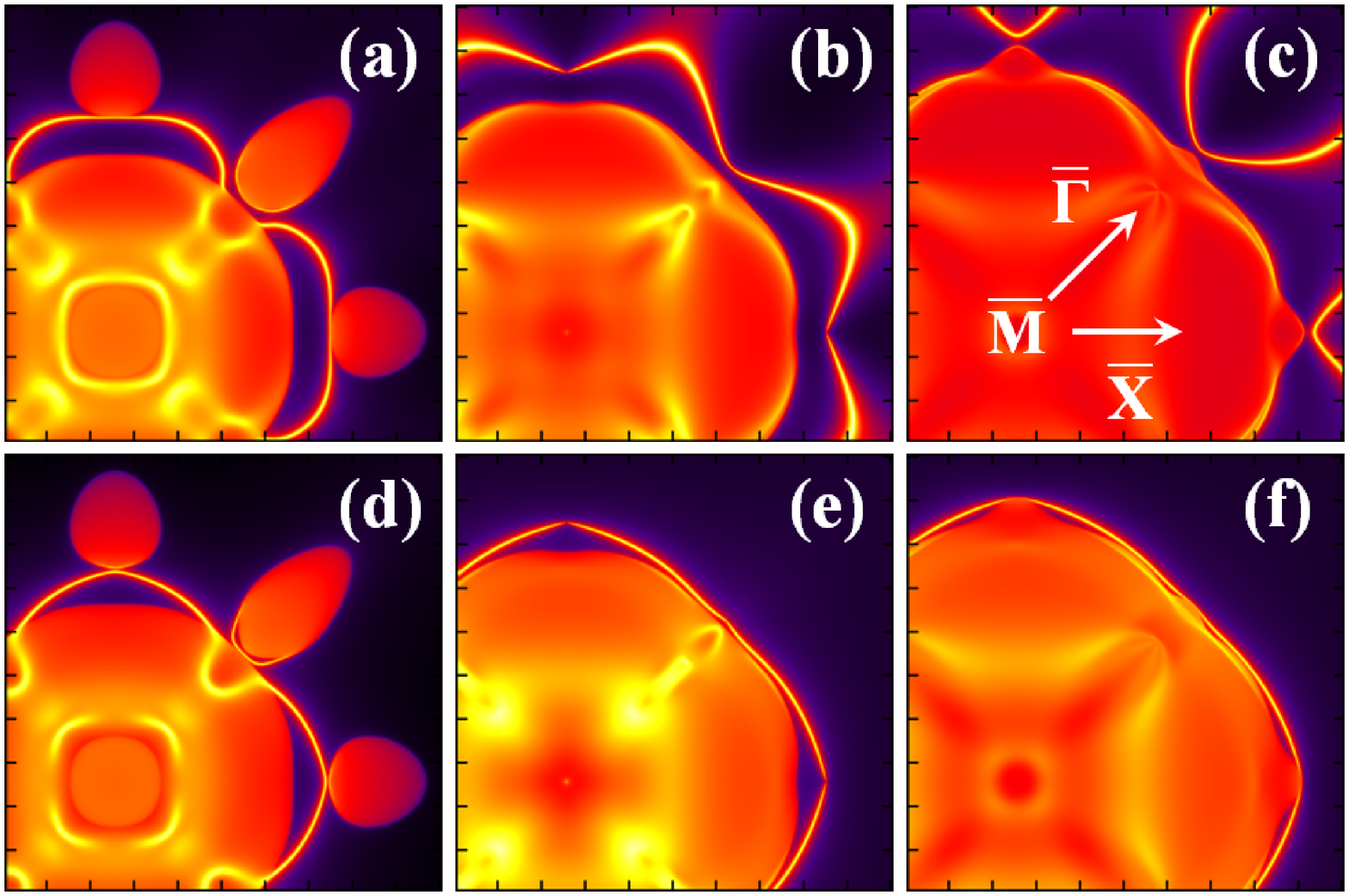}}}
\caption{Enlarged $\bar{M}$-centered (001) surface spectral 
functions for (top) Pd$_2$ and 
(bottom) Pb-Pd terminations in GGA+SOC.
Panels (a) and (d) are at $E_F$.
Panels (b) and (e) are at 0.14 eV, crossing the Dirac point along the $R-M$ line.
Panels (c) and (f) are at 0.22 eV, crossing the Dirac point along the $\Gamma-R$ line.
}
\label{s4}
\end{figure*}


\begin{thebibliography}{10}
\bibitem{armi} For a review, see
 N. P. Armitage, E. J. Mele, and A. Vishwanath,
 Weyl and Dirac semimetals in three-dimensional solids,
 Rev. Mod. Phys. {\bf 90}, 015001 (2018).

\bibitem{brad16} B. Bradlyn, J. Cano, Z. Wang, M. G. Vergniory, C. Felser, 
 R. J. Cava, and B. A. Bernevig,
 Beyond Dirac and Weyl fermions: Unconventional quasiparticles in conventional crystals,
 Science {\bf 353}, aaf5037 (2016).
 
\bibitem{zhu16} Z. Zhu, G. W. Winkler, Q. Wu, J. Li, and A. A. Soluyanov,
 Triple Point Topological Metals,
  Phys. Rev. X {\bf 6}, 031003 (2016).

\bibitem{hweng16a} H. Weng, C. Fang, Z. Fang, and X. Dai,
 Topological semimetals with triply degenerate nodal points in $\theta$-phase tantalum nitride,
 Phys. Rev. B {\bf 93}, 241202(R) (2016). 

\bibitem{hweng16b} H. Weng, C. Fang, Z. Fang, and X. Dai,
 Coexistence of Weyl fermion and massless triply degenerate nodal points,
 Phys. Rev. B {\bf 94}, 165201 (2016). 

\bibitem{yang17} H. Yang, J. Yu, S. S. P. Parkin, C. Felser, C.-X. Liu, and B.Yan,
 Prediction of Triple Point Fermions in Simple Half-Heusler Topological Insulators,
 Phys. Rev. Lett. {\bf 119}, 136401 (2017).

\bibitem{wang17} J. Wang, X. Sui, W. Shi, J. Pan, S. Zhang, F. Liu, S.-H. Wei, Q. Yan, and B. Huang,
 Prediction of Ideal Topological Semimetals with Triply Degenerate Points in the NaCu$_3$Te$_2$ Family,
 Phys. Rev. Lett. {\bf 119}, 256402 (2017).

\bibitem{ding17} B. Q. Lv,	Z.-L. Feng, Q.-N. Xu, X. Gao, J.-Z. Ma, L.-Y. Kong,	
  P. Richard, Y.-B. Huang, V. N. Strocov, C. Fang, H.-M. Weng, Y.-G. Shi, T. Qian, and H. Ding, 
 Observation of three-component fermions in the topological semimetal molybdenum phosphide,
 Nature {\bf 546}, 627 (2017).



\bibitem{phase1} Ph. Durussel and P. Feschotte,
 The binary system Pb-Pd,
 J. Alloys Comp. {\bf 236}, 195 (1996).

\bibitem{phase2} H. Okamoto,
 Pb-Pd (lead-palladium),
 J. Phase Equilibria {\bf 18}, 491 (1997).


\bibitem{kim} Y.-S. Kim and J.-M. Choi,
 X-ray diffractometric study on modification mechanism of matrixes
 for electrothermal AAS determination of volatile lead and bismuth,
 Bull. Korean Chem. Soc. {\bf 21}, 56 (2000).

\bibitem{disalvo16} Z. Cui, H. Chen, M. Zhao, and F. J. DiSalvo, 
 High-Performance Pd$_3$Pb Intermetallic Catalyst for Electrochemical Oxygen Reduction, 
 Nano Lett. {\bf 16}, 2560 (2016).

\bibitem{gunji} T. Gunji, S.-H. Noh, T. Tanabe, B. Han, C. Y. Nien, T. Ohsaka, 
and F. Matsumoto,
Enhanced Electrocatalytic Activity of Carbon-Supported Ordered Intermetallic 
Palladium–Lead (Pd$_3$Pb) Nanoparticles toward Electrooxidation of Formic Acid,
Chem. Mater. {\bf 29}, 2906 (2017).


\bibitem{ke_the} X. Ke, G. J. Kramer, and O. M. Lovvik, 
 The influence of electronic structure on hydrogen absorption in palladium alloys, 
 J. Phys.: Condens. Matter {\bf 16}, 6267 (2004).

\bibitem{kohl} H. Kohlmann, A. V. Skripov, A. V. Soloninin, and T. J. Udovic,
The anti-perovksite type hydride InPd$_3$H$_{0.89}$,
J. Solid State Chem. {\bf 183}, 2461 (2010).



\bibitem{dag94} E. Dagotto, A. Nazarenko, and M. Boninsegni,
 Flat Quasiparticle Dispersion in the 2D $t-J$ Model,
 Phys. Rev. Lett. {\bf 73}, 728 (1994).

\bibitem{imada00} M. Imada and M. Kohno,
 Superconductivity from Flat Dispersion Designed in Doped Mott Insulators,
 Phys. Rev. Lett. {\bf 84}, 143 (2000).

\bibitem{mgcni3} H. Rosner, R. Weht, M. D. Johannes, W. E. Pickett, and E. Tosatti,
 Superconductivity near Ferromagnetism in MgCNi$_3$,
 Phys. Rev. Lett. {\bf 88}, 027001 (2001). 


\bibitem{LP09} K.-W. Lee and W. E. Pickett,
 Orbital-ordering driven structural distortion in metallic SrCrO$_3$,
 Phys. Rev. B {\bf 80}, 125133 (2009).


\bibitem{gga} J. P. Perdew, K. Burke, and M. Ernzerhof,
 Generalized gradient approximation made simple,
 Phys. Rev. Lett. {\bf 77}, 3865 (1996).

\bibitem{wien2k} K. Schwarz and P. Blaha,
 Solid state calculations using WIEN2k,
 Comput. Mater. Sci. {\bf 28}, 259 (2003).


\bibitem{pbpd3} J. T. Szymanski, L. J. Cabri, and J. H. G. Laflamme,
 The crystal structure and calculated powder-diffraction, data for zvyagintsevite, Pd$_3$Pb,
 The Canadian Mineralogist {\bf 35} 773 (1997).



\bibitem{snpd3_exp1} J. R. Knight and D. W. Rhys, 
The systems palladium - indium and palladium - tin, 
J. Less-Common Metals {\bf 1}, 292 (1959).

\bibitem{snpd3_exp2} O. T. Woo, J. Rezek, and M. Schlesinger, 
X-Ray diffraction analyses of liquid-phase sintered compounds of Pd$_3$Sn and Ni$_3$Sn, 
Mater. Sci. Eng. {\bf 18}, 163 (1975).


\bibitem{marzari} A.A. Mostofi, J.R. Yates, Y.-S. Lee, I. Souza, D. Vanderbilt, 
 and N. Marzari,
 wannier90: A tool for obtaining maximally-localised Wannier functions,
 Comput. Phys. Commun.{\bf 178}, 685 (2008).

\bibitem{jan10} J. Kune\v{s}, R. Arita, P. Wissgotte, A. Toschie,
 H. Ikedaf, and K. Held,
 Wien2wannier: From linearized augmented plane waves to maximally localized Wannier functions,
 Comput. Phys. Commun. {\bf 181}, 1888 (2010).

\bibitem{wtools} Q. S. Wu, S. N. Zhang, H.-F. Song, M. Troyer, and A. A. Soluyanov,
 WannierTools: An open-source software package for novel topological materials,
 Comput. Phys. Commun. 224, 405 (2018).


\bibitem{boltz} G. K. H. Madsen and D. J. Singh,
 BoltzTraP. A code for calculating band-structure dependent quantities,
 Comput. Phys. Commun. {\bf 175}, 67 (2006).

\bibitem{optic} C. Ambrosch-Draxl and J. O. Sofo,
 Linear optical properties of solids within the full-potential linearized 
 augmented planewave method,
 Comput. Phys. Commun. {\bf 175}, 1 (2006).


\bibitem{cu3pdn} R. Yu, H. Weng, Z. Fang, X. Dai, and X. Hu,
Topological Node-Line Semimetal and Dirac Semimetal State in Antiperovskite Cu$_3$PdN,
Phys. Rev. Lett. {\bf 115}, 036807 (2015).


\bibitem{suppl} See Supplemental Material at URL
for the band structures and HWCCs of the distorted structures,
the blowup surface spectral functions.


\bibitem{zhang17} X. Zhang, Z.-M. Yu, X.-L. Sheng, H. Y. Yang, and S. A. Yang,
Coexistence of four-band nodal rings and triply degenerate nodal points in centrosymmetric metal diborides,
Phys. Rev. B {\bf 95}, 235116 (2017). 


\bibitem{green} M. P. Lopez Sancho, J. M. Lopez Sancho, J. M. L. Sancho and J. Rubio 
Highly convergent schemes for the calculation of bulk and surface Green functions,
 J. Phys. F 15, 851 (1985);
M. P. Lopez Sancho, J. M. Lopez Sancho, and J Rubio, 
Quick iterative scheme for the calculation of transfer matrices: application to Mo (100),
{\it ibid.} {\bf 14}, 1205 (1984).




\bibitem{hwcc} A. A. Soluyanov and D. Vanderbilt,
 Computing topological invariants without inversion symmetry,
 Phys. Rev. B {\bf 83}, 235401 (2011).

\bibitem{z2pack} For a brief review, 
see D. Gresch, G. Aut\'{e}s, O. V. Yazyev, M. Troyer, D. Vanderbilt, B. A. Bernevig, and A. A. Soluyanov,
Z2Pack: Numerical implementation of hybrid Wannier centers for identifying topological materials,
Phys. Rev. B {\bf 95}, 075146 (2017). 

\bibitem{fu11} L. Fu, 
 Topological Crystalline Insulators, 
 Phys. Rev. Lett. {\bf 106}, 106802 (2011).

\bibitem{kee15} Y. Chen, Y.-M. Lu, and H.-Y. Kee, 
Topological crystalline metal in orthorhombic perovskite iridates, 
Nat. Commun. {\bf 6}, 6593 (2015).


\bibitem{pbpt3_our} K.-H. Ahn, W. E. Pickett, and K.-W. Lee, (unpulished).

\bibitem{snpt3_ti} M. Kim, C.-Z. Wang, and K.-M. Ho,
 Coexistence of type-II Dirac point and weak topological phase in Pt$_3$Sn,
  Phys. Rev. B {\bf 96}, 205107 (2017).






\bibitem{boltztheory}P. B. Allen, W. E. Pickett, and H. Krakauer,
Anisotropic normal-state transport properties predicted and analyzed for
  high-T$_c$ oxide superconductors,
 Phys. Rev. B {\bf 37}, 7482 (1988).

\bibitem{nbp} K.-H. Ahn, K.-W. Lee, and W. E. Pickett,
 Spin-orbit interaction driven collective electron-hole excitations 
 in a noncentrosymmetric nodal loop Weyl semimetal,
 Phys. Rev. B {\bf 92}, 115149 (2015).

\bibitem{singh10} D. J. Singh,
 Doping-dependent thermopower of PbTe from Boltzmann transport calculations,
 Phys. Rev. B {\bf 81}, 195217 (2010).


\end{thebibliography}
\end{document}